\documentclass[a4paper]{jpconf}
\usepackage{latexsym}
\usepackage{graphicx}
\begin{document}
\title{Magnetic Order in the 2D Heavy-Fermion System CePt$_{2}$In$_{7}$ studied by $\mu^+$SR}

\author{M. M\aa nsson$^{1,2}$, K. Pr\v{s}a$^{1,3}$, Y. Sassa$^3$, P.~H.~Tobash$^4$, E. D. Bauer$^4$,\\C. Rusu$^5$, D. Andreica$^5$, O. Tjernberg$^6$, K. Sedlak$^7$, M. Grioni$^8$,\\T. Durakiewicz$^4$, and J. Sugiyama$^9$ }

\address{$^1$ Laboratory for Quantum Magnetism, $\acute{\rm E}$cole Polytechnique F$\acute{\rm e}$d$\acute{\rm e}$rale de Lausanne (EPFL), CH-1015 Lausanne, Switzerland}
\address{$^2$ Laboratory for Neutron Scattering \& Imaging, Paul Scherrer Institute, CH-5232 Villigen PSI, Switzerland}
\address{$^3$ Laboratory for Solid state physics, ETH Z\"{u}rich, CH-8093 Z\"{u}rich, Switzerland}
\address{$^4$ Los Alamos National Laboratory, Los Alamos, New Mexico 87545, USA}
\address{$^5$ Faculty of Physics, Babes-Bolyai University, 400084 Cluj-Napoca, Romania}
\address{$^6$ Materials Physics, Royal Institute of Technology KTH, S-164 40 Kista, Sweden}
\address{$^7$ Laboratory for Muon Spin Spectroscopy, Paul Scherrer Institute, CH-5232 Villigen PSI, Switzerland}
\address{$^8$ Institute of Condensed Matter Physics, \'{E}cole Polytechnique F\'{e}d\'{e}rale de Lausanne (EPFL), CH-1015 Lausanne, Switzerland}
\address{$^9$ Toyota Central Research \& Development Laboratories, Inc., 41-1 Yokomichi, Nagakute, Aichi 480-1192, Japan}

\ead{martin.mansson@epfl.ch}

\begin{abstract}
The low-temperature microscopic magnetic properties of the quasi-2D heavy-fermion compound, CePt$_{2}$In$_{7}$ are investigated by using a positive muon-spin rotation and relaxation ($\mu^{+}$SR) technique. Clear evidence for the formation of a commensurate antiferromagnetic order below $T_{\rm N} \approx$~5.40~K is presented. The magnetic order parameter is shown to fit well to a modified BSC gap-energy function in a strong-coupling scenario.
\end{abstract}

\section{Introduction}
CePt$_{2}$In$_{7}$ is a recently discovered heavy fermion material belonging to the same family as the rather famous CeIn$_3$ \cite{Mathur} and Ce$M$In$_5$ ($M=$~Co, Rh, Ir) \cite{Park,Gerber} compounds. However, the spacing
between Ce-In planes in CePt$_{2}$In$_{7}$ is drastically increased \cite{Kurenbaeva} relative to its Ce$M$In$_5$ cousins. As might be expected, the Fermi surface is more two-dimensional \cite{Altarawneh}. This compound crystallizes in the tetragonal space group $I4/mmm$ and represents a novel structure type [Pearson symbol \emph{tI}20, \emph{a}~=~4.6093 {\AA}, \emph{c}~=~21.627 {\AA}, \emph{Z}~=~2] as shown in Fig.~1(a). The magnetic susceptibility, $\chi(T)$ \cite{Bauer,apRoberts} exhibits a Curie-Weiss behavior for $T\geq$~150~K, with an effective moment $\mu_{\rm eff} =$~2.93~$\mu_{\rm B}$, which is a bit higher than what can be expected for the Ce$^{\rm 3+}$ free ion moment ($p$~=~2.54). $\chi(T)$ displays a maximum at 8~K, and an inflection at $T_{\rm N}\approx$~5~K, signaling the onset of antiferromagnetism (AF). Further, resistivity data \cite{apRoberts}, $\rho(T)$, show metallic behavior with inflections at $T_{\rm Inf1}$~=~25~K and $T_{\rm Inf2}$~=~100~K, and finally a sharp suppression below 8~K. The origin of these two high-temperature inflection points is presently unknown, but may be related to either a crystalline electric field excitation and/or the onset of coherence of the Kondo lattice \cite{apRoberts,Matsumoto}. The specific heat divided by temperature ($C/T$) exhibits a main peak at $T_{\rm N} \approx$~5.3~K associated with the AF transition \cite{Bauer,apRoberts}. Noteworthy is also the large value of $C/T \approx$~450~mJ/mol$\cdot$K$^{2}$ \cite{apRoberts}, which indicates a substantial enhancement of the effective mass, $m$*. Further, nuclear quadrupolar resonance (NQR) studies have indicated that the AF structure could be both commensurate \cite{apRoberts} as well as incommensurate \cite{Sakai}, the details of which are strongly dependent on how the polycrystalline or single crystalline samples were prepared.  This suggests that strain may affect the magnetic properties of CePt$_{2}$In$_{7}$. NQR measurements indicate that strong AF fluctuations could be present in the paramagnetic state all the way up to 4$T_{\rm N}$ \cite{apRoberts}.

\begin{figure}[t]
  \begin{center}
    \includegraphics[keepaspectratio=true,width=135 mm]{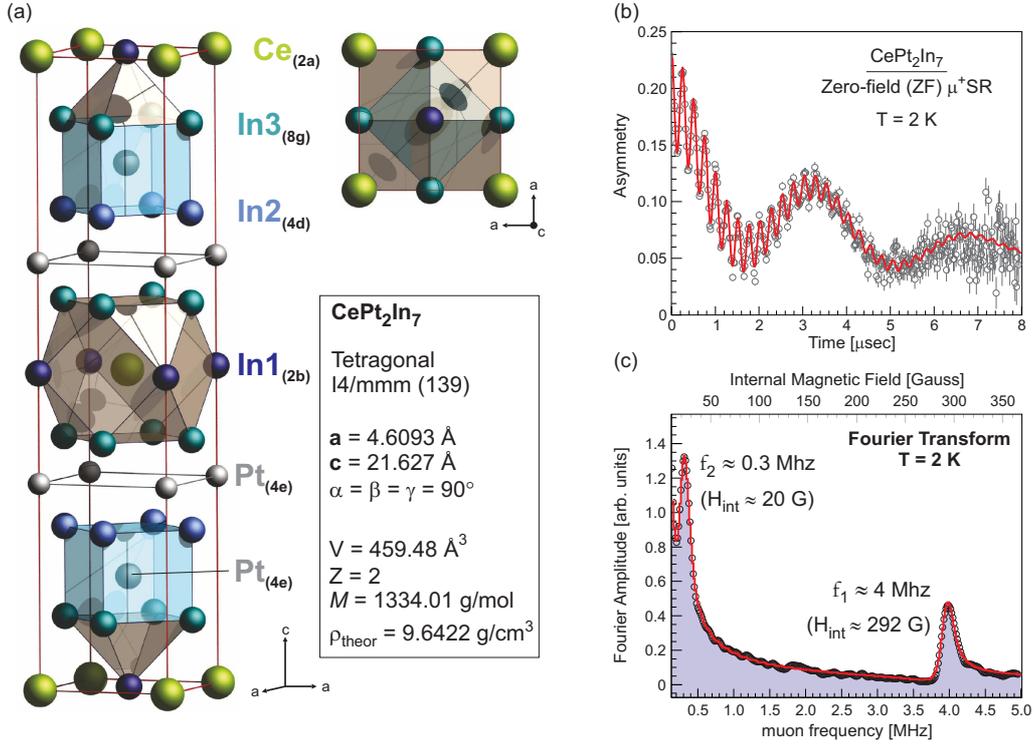}
  \end{center}
  \caption{(Color online)
  (a) Side-view and top-view of the crystal structure of CePt$_{2}$In$_{7}$.
  (b) Zero-field (ZF) $\mu^{\rm +}$SR time-spectrum acquired at $T = 2$~K.
  (c) Fourier transform of the time-spectrum at $T = 2$~K, clearly revealing the two frequencies ($f_{1}$ \& $f_{2}$).
  Solid red lines in (b) and (c) represent the fit according to Eq.~\ref{eq:ZFfit}.
    }
  \label{fig:structure}
\end{figure}

Electrical resistivity and ac-calorimetry measurements under pressure were recently carried out on single crystals of CePt$_{2}$In$_{7}$ \cite{Sidorov}.  These experiments reveal a quantum critical point at a critical pressure $P_c\approx3.2$~GPa where the AF order is completely suppressed, in which the electrical resistivity exhibits a powerlaw $T$-dependence near $P_c$. In addition, a narrow dome of superconductivity is observed around $P_c$, suggesting that critical AF fluctuations may mediate the Cooper pairing. In contrast, a broad superconducting dome is found in poylcrystalline samples \cite{Bauer2}, indicating that strain at the grain boundaries also affects the superconductivity.  NMR/NQR measurements under pressure on single crystals \cite{Sakai2} reveal that a localized-itinerant crossover of the Ce-4$f$ electron occurs within the AF state at $P^{*}\approx2.5$~GPa, approximately the pressure where superconductivity first emerges.

\section{Experimental Details}
Approximately 1.5 grams of polycrystalline CePt$_{2}$In$_{7}$ sample in the ratio 1:2:7.5 were prepared by arc-melting on a Cu hearth under UHP Ar atmosphere with a Zr getter. The resulting compound was subsequently wrapped in Ta foil and annealed under vacuum at 500$^{\circ}$C for 2 weeks to stabilize the 1:2:7 phase. The sample preparation procedure is described in greater detail elsewhere \cite{Bauer,Tobash}. For the $\mu^+$SR experiment the sample was placed in a small envelope made of $50~\mu$m thin Al-coated Mylar tape and then attached to a low-background fork-type sample holder. In order to make certain that the muons stopped primarily inside the sample, we ensured that the side facing the muon beamline was covered only by a single layer of Mylar tape. Subsequently, $\mu^+$SR spectra were recorded at the Swiss Muon Source (S$\mu$S), Paul Scherrer Institut, Villigen, Switzerland. By using the DOLLY spectrometer at the surface muon beamline $\mu$E1, zero-field (ZF) and weak transverse-field (wTF) data were collected for 0.25~K~$\leq{}T\leq$~20~K. The experimental setup and techniques were described in detail elsewhere \cite{muSR_book}.

\section{Results and Discussion}
$\mu^+$SR spectra recorded in ZF at $T=2$~K [see Fig.~1(b-c)] show clear oscillations indicating that the muon-spins are precessing around spontaneous internal magnetic fields created, at the muon site, by the AF ordering. The fast Fourier transform (FFT) of the raw time spectrum clearly shows that the oscillating signal is composed of two different frequencies $f_{1}\approx4$~MHz and $f_{2}\approx0.3$~MHz. Accordingly, the ZF muSR spectra recorded at temperatures well below $T_{\rm N}$ were well fitted by the combination of two exponentially damped cosine oscillations corresponding to an average of 2/3 of the muons that "see" a magnetic field perpendicular to the muon's initial polarization, a slowly relaxing non-oscillatory signal corresponding to the remaining 1/3 of the muons for which $H_{\rm int}$ is parallel to their initial polarization, and a small background (BG) signal from a minor fraction of the muons stopping in the sample holder:
\begin{eqnarray}
 A_0 \, P_{\rm ZF}(t) &=&
A_{\rm 1}\cos(2\pi\cdot f_{\rm 1}t+\phi_1)\exp(-\lambda_{\rm 1} t)
\cr
 &+& A_{\rm 2}\cos(2\pi\cdot f_{\rm 2}t+\phi_2)\exp(-\lambda_{\rm 2} t)
\cr
 &+& A_{\rm tail}\exp(-\lambda_{\rm tail} t)
 \cr
 &+& A_{\rm BG}\exp(-\lambda_{\rm BG} t)
 \cr
 &+& A_{\rm KT}~G^{\rm SGKT}(\Delta_{\rm KT},t),
\label{eq:ZFfit}
\end{eqnarray}
Here, $A_0$ is the initial ($t$~=~0) asymmetry, $P_{\rm ZF}(t)$ is the muon-spin polarization function in ZF, $f_{\rm 1}$ and $f_{\rm 2}$ are the two muon Larmor frequencies related to the internal fields as $f_{\rm i}=(\gamma_{\mu}/2\pi)\times B_{\rm int}$, $\phi_1$ \& $\phi_2$ are the initial phases and $\lambda_{\rm 1}$ \& $\lambda_{\rm 2}$ are the exponential relaxation rates of the precessing signals. Further, $\lambda_{\rm tail}$ is the exponential relaxation rate of the tail signal, while $\Delta_{\rm KT}$ is the static field distribution width of the isotropic dipolar field from the frozen (nuclear) magnetic moments. Finally, $A_{\rm 1}$, $A_{\rm 2}$, $A_{\rm tail}$ and $A_{\rm BG}$ are the asymmetries (i.e. volume fraction) of the four individual signals. As seen from Eq.~\ref{eq:ZFfit}, also a fifth component is added with a static Gaussian Kubo-Toyabe (KT) function \cite{KT} [$G^{\rm SGKT}(\Delta_{\rm KT},t)$] that describes the presence of randomly oriented (frozen) spin system with a static moment. This last component only becomes important in the cross-over region close to $T_{\rm N}$ where the fit function evolves from Eq.~\ref{eq:ZFfit} to a simple Kubo-Toyabe function as the transition from AF to PM order occurs.

\begin{figure}[t]
  \begin{center}
    \includegraphics[keepaspectratio=true,width=135 mm]{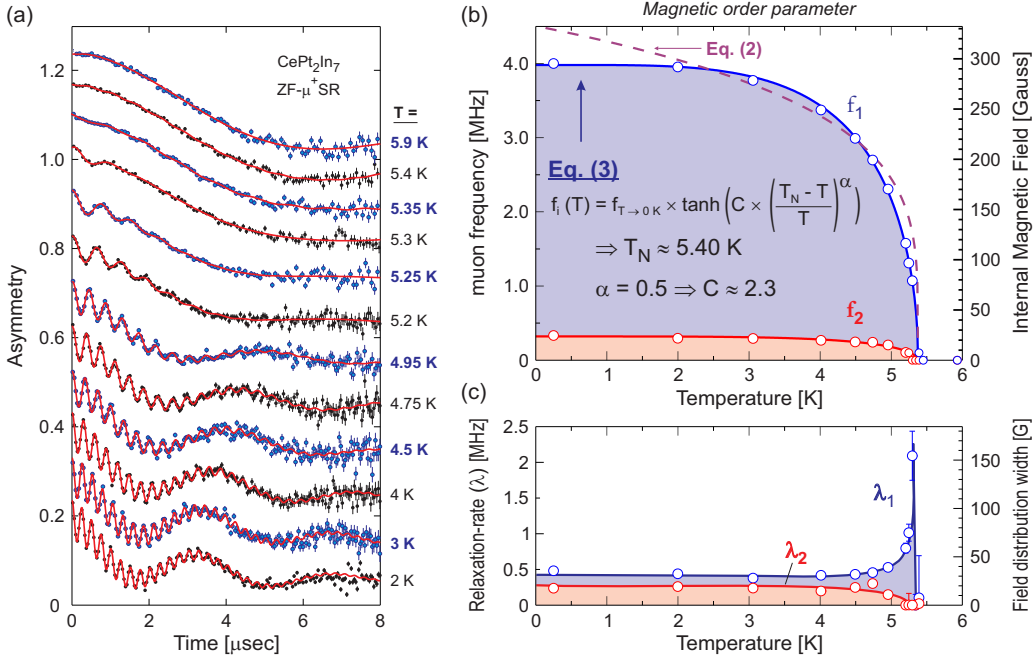}
  \end{center}
  \caption{(Color online)
  (a) Temperature dependent ZF $\mu^+$SR spectra where the solid (red) lines represent the fit result using Eq.~(\ref{eq:ZFfit}). Each spectrum is offset by 0.1 in asymmetry for clarity of display.
  (b) Magnetic order parameter fitted to Eq.~\ref{eq:MF} (dashed line) and Eq.~\ref{eq:BCS} (solid lines), respectively.
  (c) Temperature-dependence of the relaxation rates, $\lambda_{i}$. Solid lines in (c) are only guides to the eye.
    }
  \label{fig:results}
\end{figure}

As seen from Fig.~\ref{fig:results}(a) and from the temperature dependencies of $f_{\rm 1}$ and $f_{\rm 2}$ shown in Fig.~\ref{fig:results}(b), as $T$ increases the oscillation frequencies are gradually suppressed and finally disappear around $T^{\mu}_{N} \approx 5.40$~K. The magnetic order parameter shown in Fig.~\ref{fig:results}(b) could only be rather poorly fitted by a basic mean-field theory approach:
\begin{eqnarray}
 f_{\rm i} &=& f_{\rm T\rightarrow 0 K} \cdot \left(\frac{T_{\rm N}-T}{T_{\rm N}}\right)^{\beta}
\label{eq:MF}
\end{eqnarray}
With $T_{\rm N} = 5.40$~K the critical exponent $\beta \approx$~0.25, which is in the intermediate range between the predictions for a 2D and 3D Ising model ($\beta =$~0.125 and 0.3125, respectively) \cite{Collins,Stanley}. Such result not only contradict the fact that CePt$_{2}$In$_{7}$ is expected to have a strong 2D character, but also generate a clearly inadequate fit for both the high and low-temperature regime [see dashed line in Fig.~2(b)]. The metallic conductivity of this material instead imply that the magnetic order parameter [Fig.~2(b)] could be fitted by the BCS gap-energy function that is commonly used for a spin-density wave (SDW) order parameter \cite{Gruner}:
\begin{eqnarray}
 f_{\rm i} &=& f_{\rm T\rightarrow 0 K} \cdot \tanh\left[C \cdot \left(\frac{T_{\rm N}-T}{T}\right)^{\alpha}\right]
\label{eq:BCS}
\end{eqnarray}
Here $C = 1.74$ and $\alpha = 1/2$ are for the conventional weak-coupling BSC scenario while a higher value of $C$ is a sign for the presence of strong-coupling effects \cite{Tinkham,Betouras}. Fitting the magnetic order parameter to Eq.~\ref{eq:BCS} in the weak-coupling scenario, yields a very poor result. However, when allowing also $C$ to be fitted a very good result if obtained for $T^{\mu}_{N} \approx 5.40$~K, $C \approx 2.3$ and $\alpha = 1/2$ (\emph{fixed}), as shown in Fig.~2(b) (solid lines). This is a clear indication that a strong-coupling scenario is more appropriate for this compound, as could be expected from the reported high effective mass, $m$* resulting from the interactions of the conduction electrons with the magnetic Ce ions \cite{Fisk}.

The initial phases of the oscillations, $\phi_1$ and $\phi_2$, are equal and close to zero in the whole temperature range. Such fact together with a rather symmetric field-distribution shown in the FFT spectrum [Fig.~1(c)], suggest that the AF order is commensurate to the crystallographic lattice, as has been partly proposed by NQR \cite{apRoberts}. However, it should be emphasized that neutron diffraction would be the appropriate tool, more suitable than both NQR and $\mu^+$SR, for robustly determining the detailed spin-structure of CePt$_{2}$In$_{7}$. Further, that $f_{1}$ and $f_{2}$ show the same temperature dependence and are equally well fitted to Eq.~\ref{eq:BCS}, clearly suggests that the multiple frequencies are not caused by the coexistence of two different phases in the sample but a direct result of two magnetically inequivalent interatomic muon stopping sites within the crystal lattice.

Figure 2(c) shows the temperature dependence of the relaxation rate (field distribution width) for the fast oscillation ($\lambda_{1}$), which display a critical behavior close to the transition. This is caused by that the AF fluctuations become stronger and stronger with increasing temperature, until the static AF order is completely destroyed at $T_{\rm N}$. For the slow oscillation, $\lambda_{2}$ appears to display a different temperature dependence close to the transition. However, this is most likely merely a fitting-effect caused by a problem to disentangle the slow oscillation from the evolving KT function close to $T_{\rm N}$.

\section{Conclusions}
By the use of a positive muon-spin rotation and relaxation ($\mu^{+}$SR) technique we find clear evidence for the formation of a commensurate antiferromagnetic order in CePt$_{2}$In$_{7}$ below $T_{\rm N} \approx$~5.40~K. The magnetic order parameter can be well fitted to a modified BSC gap-energy function in a strong-coupling scenario.

\section{Acknowledgments}
This work was performed at the Swiss Muon Source (S$\mu$S), Paul Scherrer Institut (PSI), Villigen, Switzerland and we are thankful to Robert~Scheuermann, Bastian Wojek and Matthias Elender for assistance with the $\mu^+$SR experiments. This work is partially supported by the Swiss National Science Foundation (SNSF), by Grant-in-Aid for Scientific Research on Innovative Areas "Ultra Slow Muon" (No.~23108003) of the Ministry of Education, Culture, Sports, Science, and Technology, Japan, MEXT KAKENHI Grant No. 23108003, JSPS KAKENHI Grant No. 26286084, and the Swedish Research Council (VR). Work at LANL was performed under the auspices of the U.S. DOE, Office of Basic Energy Sciences, Division of Materials Sciences and Engineering. D.A. acknowledges financial support from the Romanian UEFISCDI project No. PN-II-ID-PCE-2011-3-0583 (85/2011). All images involving crystal structure were made with DIAMOND and the $\mu^+$SR data was fitted using \texttt{musrfit} \cite{musrfit}.

\section*{References}

\end{document}